\documentclass[%
 reprint,
 amsmath,amssymb,
 aps,
 pra,
]{revtex4-1}

\usepackage{graphicx}
\usepackage{dcolumn}
\usepackage{bm}
\usepackage{float}
\usepackage{xcolor}
\usepackage{mathtools}
\usepackage{etoolbox}
\usepackage{physics}
\usepackage{amsmath}
\usepackage{xparse}
\usepackage[titletoc]{appendix}

\usepackage{soul}
\usepackage{alphabeta}

\def \beq {\begin{equation}}
\def \eeq {\end{equation}}

\begin{document}

\title {Non-Markovianity in the time evolution of open quantum systems assessed by means of quantum state distance} 

\author{G. Mouloudakis$^{1,2}$}
 \email{gmouloudakis@physics.uoc.gr}

\author{I. Stergou$^{1}$}

\author{P. Lambropoulos$^{1,2}$}

\affiliation{${^1}$Department of Physics, University of Crete, P.O. Box 2208, GR-71003 Heraklion, Crete, Greece
\\
${^2}$Institute of Electronic Structure and Laser, FORTH, P.O.Box 1527, GR-71110 Heraklion, Greece}

\date{\today}

\begin{abstract}

We provide a quantitative evaluation of non-Markovianity (NM) for an XX chain of interacting qubits with one end coupled to a reservoir. The NM of several non-Markovian spectral densities is assessed in terms of various quantum state distance (QSD) measures. Our approach is based on the construction of the density matrix of the open chain, without the necessity of a master equation. For the quantification of NM we calculate the dynamics of the QSD measures between the Markovian-damped and various types of non-Markovian-damped cases. Since in the literature several QSD measures, appear in forms that imply trace preserving density matrices, we introduced appropriate modifications so as to render them applicable to the case of decaying traces. The results produce remarkable consistency between the various QSD measures. They also reveal a subtle and potentially useful interplay between qubit-qubit interaction and non-Markovian damping. Our calculations have also uncovered a surprisingly dramatic slowing-down of dissipation by the squared Lorentzian reservoir.

\end{abstract}

\maketitle

\section{Introduction}

The dynamics of open quantum systems coupled to non-Markovian reservoirs is at the forefront of current interest, as it is relevant to practical as well as to fundamental problems. In that endeavor, one dimensional many-body systems such as quantum spin chains arise in many contexts, as they offer valuable and versatile ground for quantitative studies. Their relevance extends from quantum information theory to  condensed matter physics, owing to their versatility as basic resources for the implementation of solid-state devices for quantum computing and quantum communication tasks \cite{ref1}. Among these tasks, faithful quantum state transfer \cite{ref2,ref3,ref4,ref5,ref6,ref7,ref8,ref9} and long-distance entanglement \cite{ref10,ref11,ref12,ref13,ref14,ref15} have been investigated for a variety of spin chain configurations,  over the last 20 years or so, with  research on these fields continuing at an undiminished pace. 

In that general context, the dynamics of an arbitrary length spin XX chain driven by non-Markovian reservoirs at both ends was studied in a recent paper by two of us \cite{ref16}. Our contribution in that paper was the detailed quantitative study of reservoirs, with Lorentzian as well as Ohmic spectral densities. Through a formulation in terms of the wave-function, combined with Laplace transform, we were able to obtain results for chains of arbitrary length, i.e. number of sites, and study dynamical as well as state transfer aspects of the system. Guided by the behaviour of the excitation dynamics, we assessed the qualitative differences of the system from that driven by a Markovian reservoir which is known to entail monotonic decay. That assessment, indirect as it was, can only be considered qualitative, although it did provide a glimpse of the similarities between these two non-Markovian cases. Yet, there are a number of technical measures aiming at a quantitative assessment of the difference between the states of a quantum system, as it evolves under coupling to different external causes; in this case reservoirs. The difference between the evolution under a given reservoir from that under a Markovian one, should therefore provide a quantitative measure for what, following standard terminology \cite{ref17,ref18}, we shall refer to as non-Markovianity” (NM). The term degree of Markovianity (DM) can also be found in the literature, meant to indicate the similarity  to the evolution under a Markovian reservoir. Since  the terms environment and reservoir appear in the literature as synonyms, we are using them interchangeably in this paper. Although we have modelled the reservoirs in terms of an infinite collection of bosons, less common but essentially equivalent possible modelling in terms of fermions can be found in the literature \cite{ref19}. By its nature, our system consists of fermions which could also be handled via the Jordan-Wigner transformation \cite{ref20,ref21}. The route we have chosen has been dictated by simplicity and usefulness in state transfer problems. 

Having examined much of the existing literature on the issue, we have come to the realization that our scheme provides excellent territory for the quantitative study of non-Markovianity. The relevant literature is vast, ranging from general formal considerations \cite{ref22,ref23,ref24,ref25,ref26,ref27} to examples in small specific systems \cite{ref28,ref29,ref30,ref31}, including methods for the estimation of NM using machine learning tools \cite{ref32}. The advantage of our system stems from the combination of a realistic arrangement of an XX chain of mutually interacting qubits, with a reservoir of arbitrary spectral density coupled to one end. In our previous paper \cite{ref16}, we had considered a chain coupled to reservoirs at both ends. Although the present quantitative study could as easily address that arrangement, since our objective is the comparative study of non-Markovian reservoirs, we have chosen to consider the same system with only one end qubit coupled to a variety of reservoirs, so as to focus on the role of the reservoir. We calculate first the time evolution of the system under each reservoir, one of which is Markovian. In the second step, we calculate the difference of the state of the system evolved under a given non-Markovian reservoir from the state evolved under the Markovian one. The time evolution of that difference is what we refer to as the NM of the given reservoir. 

We have chosen to address the issue in terms of perhaps the most direct notion, namely the quantum state distance (QSD). A variety of measures appropriate as tools for our task have been proposed over the years \cite{ref33,ref34}. Their objective is the quantification of the difference between two different quantum states of a system, which is of fundamental importance in quantum information processing \cite{ref35,ref36}, state transfer being a case in point \cite{ref37}. The most common quantity pertaining to that issue is the Fidelity (F), for which there are various expressions in the literature \cite{ref38,ref39,ref40}. It is meant to characterize the similarity or difference between the density matrices representing the two states. Obviously it is related to the notion of distance between two quantum states, which explains why the formulation of some distance measures involve one of the F expressions. There is no unique expression or value for the F, which is also the case for the QSD.  In some sense, it may be matter of taste or perhaps usefulness in a particular context which value of F or QSD is adopted.

Our stated objective then is the quantitative evaluation and calibration of non-Markovianity of several non-Markovian reservoirs by means of QSD. In view of the diversity of expressions for the QSD, inevitably an additional component has been injected into our task, namely, the comparative analysis of various QSD measures. Specifically, since the NM does depend on the particular QSD measure, it is important to know whether and to what extend the classification of various reservoirs in terms of NM is independent of the QSD measure employed. And this completes the road map of our present project.

In the next session, we review various QSD measures, their properties and the modifications necessary in order to accommodate issues pertaining to open systems. In section III we describe the theoretical approach to our problem in terms of the time-dependent Schr{\"o}dinger’s equations of motion in Laplace space, from which one can obtain even closed form solutions for the judiciously transformed amplitudes, for chains of arbitrary  length, as well as arbitrary initial conditions, within the single-excitation subspace. Using the amplitudes of the chain sites, we construct the density matrices necessary for the calculation of the various QSD measures. In section IV we present the results of our study while in section V we provide an overview as well as some concluding remarks related to our work.

\section{Brief summary and necessary modifications of QSD measures}

Before embarking on the discussion of QSD measures, a significant clarification related to our approach and treatment is necessary. As discussed in the next section, the time evolution of the system is formulated in terms of the time-dependent Schr{\"o}dinger equation, from which we obtain the amplitudes of the wave-function of the system as a function of time, after having eliminated the degrees of freedom of the reservoir. The time-dependent amplitudes are then combined to obtain the reduced density operator of the system which is needed in all expressions of QSD. The advantage of this approach is that it applicable for any form of reservoir, sidestepping thus the need for a master equation which is not generally available, especially for non-Markovian resevoirs. There is, however, a price paid for this advantage. The resulting density operator describes only the excitation, because the populations of the ground states have been discarded in the process of eliminating the continuum of the reservoir. In other words, we retain only one part of the entire Hilbert space of the compound system. Inclusion of those infinite in number terms (see Eqn. (18c) in the following section) is computationally impossible, as it would entail keeping track of the degrees of freedom of the reservoir. The implication for the system density operator is that its trace eventually decays to zero. The situation is analogous to the treatment of the decay of a discrete state coupled to a continuum \cite{ref41,ref42}, except that here we deal with the excitation of a chain. Thus in a real sense, the reduced density operator describes the time evolution of the excitation which as expected, in the long time limit disappears into the reservoir. Our study aims at examining the role of the form of a reservoir in that time evolution. 

In books and papers, more often than not, expressions for F or QSD are assumed, either tacitly or explicitly, to involve density operators of trace equal to unity. However, since in our approach, the trace of the reduced density operator decays, the expression for the QSD needs to be modified accordingly.  

To illustrate the issue, let us consider the most common and straightforward QSD measure, referred to as the trace distance. It is defined by
\beq
D_T \equiv \frac{1}{2}  \Tr \abs{ρ-σ}
\eeq
where $\abs{A}$, for a matrix $A$, stands for $\sqrt{A^{\dagger}A}$ which stands for the positive definite square root of the matrix under the radical symbol, with $A^{\dagger}$ being the Hermitian adjoint of $A$. Although slight variations of the definition of the square root of a matrix can be found in the literature, we adopt the one most commonly found in the literature on quantum information \cite{ref43}. Applying the above definition to the expression for the trace distance, we obtain

\beq
D_T = \frac{1}{2}  \Tr \sqrt{ \left( ρ^{\dagger} ρ + σ^{\dagger} σ - ρ^{\dagger} σ - σ^{\dagger} ρ \right)}
\label{D_T_extended}
\eeq

Since the density operators under the radical are Hermitian, the expression for the trace distance shown in Eqn. (\ref{D_T_extended}) reduces to

\beq
D_T = \frac{1}{2}  \Tr \left( ρ-σ \right)
\eeq
which is one of the possible square roots. It so happens that the trace of that root is identically zero for matrices normalized to unity. However, in our case the traces of the density operators are time-dependent, decaying to zero in the long time limit. Consequently the trace distance is also time-dependent in a fashion depending on the relevant reservoirs, as calculated through Eqn. (\ref{D_T_extended}).

Let us consider now another expression for QSD, namely the Hellinger measure $D_H$, defined as

\beq
D_H^2 \equiv \Tr \left( \sqrt{ρ} - \sqrt{σ} \right)^2
\eeq
which upon expanding the square becomes
\beq
D_H^2 = \Tr ρ + \Tr σ - 2\Tr \left( \sqrt{ρ} \sqrt{σ} \right)
\label{Hellinger_squared}
\eeq
where the invariance of the trace under cyclic permutation of the factors has been used. If the density operators are and remain normalized to unity, Eqn. (\ref{Hellinger_squared}) reduces to $D_H^2 = 2 \left[ 1- \Tr \left( \sqrt{ρ} \sqrt{σ} \right)  \right]$, leading to the expression
\beq
D_H = \sqrt{2} \left[ 1- \Tr \left( \sqrt{ρ} \sqrt{σ} \right)  \right]^{1/2}
\label{Hellinger_closed}
\eeq
for the Hellinger distance measure, which typically is the expression cited for this QSD measure. However, in order to account for our case, following Eqn. (\ref{Hellinger_squared}), we have:

\beq
D_H = \left[  \Tr ρ + \Tr σ - 2\Tr \left( \sqrt{ρ} \sqrt{σ} \right) \right]^{1/2}
\label{Hellinger_open}
\eeq

If the density operators tend to zero as $ t \rightarrow \infty$, in that limit, Eqn. (\ref{Hellinger_open}) correctly yields zero for the distance between the two states, as they both decay. On the other hand, Eqn. (\ref{Hellinger_closed}) would lead to the value $\sqrt{2}$ which on physical grounds, is at best problematic, as it does not account for the decay of the traces of the density operators.

There is one more measure that we employ in our calculations. It is known as the Bures distance measure, usually denoted by $D_B$. Since it is typically defined in terms of one of the expressions for F, we list below the three most common expressions, labelled for our convenience $F_1$, $F_2$  and $F_3$. They are:

\begin{subequations}
\beq
F_1 \left( ρ, σ \right) = \left( \Tr \sqrt{ \sqrt{ρ} σ \sqrt{ρ}}  \right)^2
\eeq

\beq
F_2 \left( ρ, σ \right) = \Tr \sqrt{\sqrt{ρ} σ \sqrt{ρ}} = \sqrt{ F_1 \left( ρ, σ \right)}
\label{F_2}
\eeq

\beq
F_3 \left( ρ, σ \right) = \Tr \left( ρ σ \right)
\eeq
\end{subequations}

The Bures QSD is usually defined as $D_B^2 \left( ρ, σ \right) \equiv 2 \left[ 1 - \sqrt{ F_1 \left( ρ, σ \right)} \right]$, which, in view of Eqn. (\ref{F_2}) above, can be simply written as

\beq
D_B^2 \left( ρ, σ \right) = 2 \left[ 1 -  F_2 \left( ρ, σ \right) \right]
\label{Bures_QSD}
\eeq
from which we obtain

\beq
D_B \left( ρ, σ \right) = \sqrt{2} \left[  1 - \Tr \sqrt{\sqrt{ρ} σ \sqrt{ρ}} \right]^{1/2}
\eeq
Again, this expression is valid as long as the traces of both density operators remain equal to one, because if in any distance measure we set $ρ=σ$, we should obtain zero. This condition is satisfied as long as  $\Tr ρ = \Tr σ =1$ for all times. But if the traces decay to zero, for the case $ρ = σ$ we would obtain the nonphysical value $\sqrt{2}$. This contradiction is amended if the above expression is modified as shown in the following equation: 

\beq
D_B \left( ρ, σ \right) = \sqrt{2} \left[ \frac{1}{2}\left(\Tr ρ+ \Tr σ \right) - \Tr \sqrt{\sqrt{ρ} σ \sqrt{ρ}} \right]^{1/2}
\label{Bures_open}
\eeq
This is the expression we are using for the Bures QSD in this paper. In the long time limit and in the presence of dissipation, the modified Bures distance tends to zero as it should. Part of our investigation addresses the rate with which the various QSD measures tend to their final value. We are in addition interested in the consistency of the character of NM obtained by different measures. For whatever it is worth, it seems to us that one could as well define a QSD measure through an extension of Eqn. (\ref{Bures_QSD}), by inserting any of the expressions for F.

\section{Theory}

\begin{figure*}[t] 
	\centering
	\includegraphics[width=17.8cm]{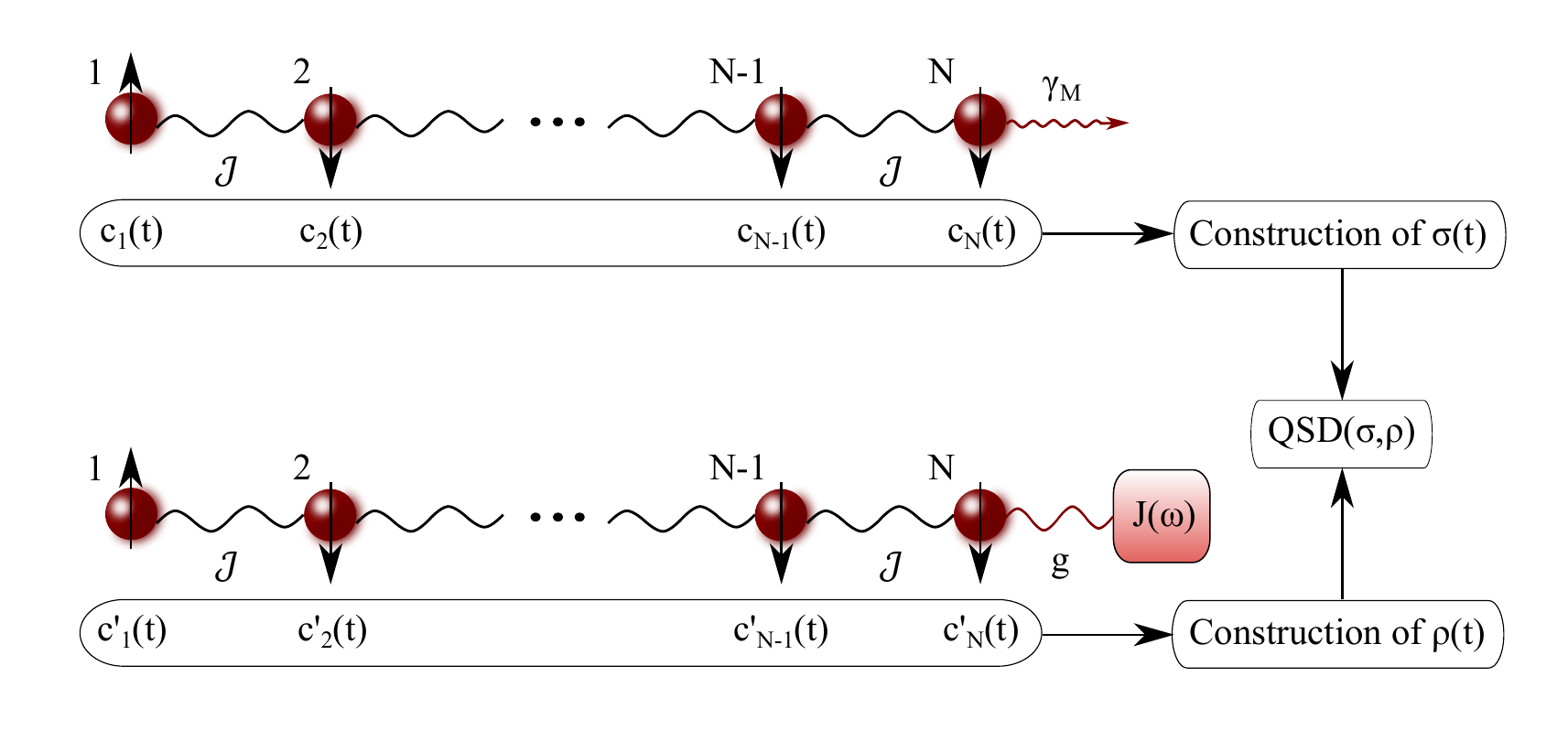}
		\caption[Fig1]{Schematic representation of the system at study and our methodology. A chain of an arbitrary number of coupled qubits is interacting with an external reservoir at its one end. The system is studied for two different cases of external reservoirs, namely for a Markovian and a non-Markovian reservoir with spectral density $J(ω)$. Through our methodology we calculate the qubit amplitudes in the single-excitation subspace via which we construct the density matrix of the chain for each case. The QSD between the two density matrices is then calculated using various measures, allowing us to assess the degree of non-Markovianity of the system for different types of spectral densities $J(ω)$.}
		\label{Fig1}
\end{figure*}

As discussed in section II, all expressions for QSD measures involve the density matrices $ρ$ and $σ$ of the quantum states whose distance is to be calculated. In general, the temporal evolution of the density matrix of a system coupled to an external environment is given by the solution of a master equation, obtained after tracing out the environmental degrees of freedom from the density matrix of the compound system (system + environment) \cite{ref41}. Despite the ongoing progress in the derivation of closed form expressions for master equations of quantum systems coupled to non-Markovian reservoirs \cite{ref44,ref45,ref46,ref47,ref48}, such expressions are usually obtained for special cases of environmental spectral densities under certain approximations and often are too complicated to be handled even numerically. Consequently, an attempt to quantify the degree of non-Markovianity in systems coupled to environments with a variety of possible spectral densities using numerical solutions of the respective master equations would be pointless -if possible at all- for certain forms of environmental spectral densities, such as those studied in this work.

To address this difficulty, we have developed  an approach enabling the calculation of the amplitudes of the wavefunction of the open system, from which we then construct its density matrix. This formulation has been described in detail in a previous paper \cite{ref16}, where it was used to calculate aspects of the dynamics of a XX chain boundary-driven by non-Markovian environments. As demonstrated below, one of the advantages of this approach is its applicability to any possible form of environmental spectral density, as well as an arbitrary number of qubits in the chain.

The system under consideration consists of $N$ identical qubits in a one-dimensional configuration coupled to each other via a nearest-neighbor coupling $\mathcal{J}$. The $N^{th}$ qubit of the chain is coupled, via a coupling strength $g$, to an external environment of an arbitrary spectral density $J(ω)$. Our goal is to assess the degree of non-Markovianity of that system by means of the QSD between its density matrix and the density matrix of the same system of $N$ qubits, with the $N^{th}$ qubit coupled to a Markovian environment. The Markovian environment is known to induce a shift of the energy of the last qubit, as well as a decay \cite{ref41}. We neglect the shift, as it is of no significance in the issue of dissipation and denote the decay by $\gamma_M$. The problem is studied using several QSD measures, for various forms of non-Markovian spectral densities $J(ω)$. Hereafter, we denote the density matrix of the Markovian open system by $σ(t)$ and the density matrix of the non-Markovian open system by $ρ(t)$. A schematic presentation of the systems under study and our methodology is depicted in Fig. 1. 

The Hamiltonian of the compound system $\mathcal{H} = \mathcal{H}_S + \mathcal{H}_E + \mathcal{H}_I $, consists of three parts; namely, the Hamiltonian of the XX chain $\mathcal{H}_S$, the Hamiltonian of the environment coupled to $N^{th}$ (last) qubit of the chain $\mathcal{H}_E$, and the chain-environment interaction Hamiltonian $\mathcal{H}_I$, given by the relations $(\hbar = 1)$:

\begin{subequations}
\beq
\begin{split}
\mathcal{H}_S   =  & \omega_e \sum_{i=1}^N \ket{e}_i \prescript{}{i}{\bra{e}}  + \omega_g \sum_{i=1}^N \ket{g}_i \prescript{}{i}{\bra{g}}  \\ 
& + \sum_{i=1}^{N-1} \frac{\mathcal{J}}{2} \left( \sigma_i^{+} \sigma_{i+1}^{-} + \sigma_i^{-} \sigma_{i+1}^{+} \right),
\end{split}
\eeq

\beq
\mathcal{H}_E = \sum_\lambda \omega_{\lambda} {a^{E}_{\lambda}}^{\dagger} a_{\lambda}^{E} ,
\eeq

\beq
\mathcal{H}_I =  \sum_\lambda g ( \omega_{\lambda} ) \left( a_{\lambda}^{E} \sigma_N^{+} + {a_{\lambda}^{E}}^{\dagger} \sigma_N^{-} \right),
\eeq
\end{subequations}
where $\omega_g$ and $\omega_e$ are, respectively, the energies of the ground and excited state of each spin (all of which are assumed identical),  $\omega_{\lambda}$  is the energy of the $\lambda$-mode  of the environment consisting of an infinite number of bosons, $a_{\lambda}^{E}$  and ${a^{E}_{\lambda}}^{\dagger}$, are the bosonic annihilation and creation operators of the environment,  and $\sigma_i^{+}= \ket{e}_i \prescript{}{i}{\bra{g}}$ and $\sigma_i^{-}= \ket{g}_i \prescript{}{i}{\bra{e}}$, $i=1, \dots, N$ are the qubit raising and lowering operators, respectively.

The wavefunction of the compound system in the single-excitation space is expressed as:

\beq
\ket{\Psi (t)} = \sum_{i=1}^N c_i(t) \ket{\psi_i}  + \sum_{\lambda} c_{\lambda}^{E}(t) \ket{\psi_{\lambda}^{E}},
\eeq
where,

\begin{subequations}
\beq
\ket{\psi_i} \equiv \ket{g}_1 \ket{g}_2 \dots \ket{g}_{i-1} \ket{e}_i \ket{g}_{i+1} \dots \ket{g}_N \ket{0}_{E},
\eeq
and
\beq
\ket{\psi_{\lambda}^{E}} \equiv \ket{g}_1 \ket{g}_2 \dots \ket{g}_N \ket{0 0 \dots 0 1_{\lambda} 0 \dots 0 0}_{E}.
\eeq
\end{subequations}

\subsection{Construction of $σ(t)$ (Markovian system)}
As already noted above, a Markovian environment (i.e. an environment with slowly varying spectral density) interacting with a two-level quantum system (qubit) within the Born approximation, induces a shift as well as a decay in the energy of the latter. Formally, this result is obtained by tracing  the environmental degrees of freedom out of the density matrix of the compound system, a procedure found in many quantum optics and open quantum systems textbooks \cite{ref41, ref49}.  In our problem, the Markovian environment will only affect  the energy of the last qubit ($N^{th}$ qubit) of the chain. Neglecting the shift and keeping the decay rate $\gamma_M$, the time dependent Schr\"{o}dinger equation (TDSE) for the chain amplitudes reads:

\begin{subequations}
\beq
\begin{split}
i \frac{dc_i(t)}{dt} = & \left[ \omega_e + (N-1) \omega_g \right] c_i(t) + \frac{\mathcal{J}}{2} \left[ c_{i-1} (t) + c_{i+1} (t) \right], \\
& \quad i = 1, \dots, N-1,
\end{split}
\label{Ampl_j_Markovian}
\eeq
\beq
i \frac{dc_N(t)}{dt}= \left[ \omega_e + (N-1) \omega_g -i γ_M \right] c_N(t) + \frac{\mathcal{J}}{2} c_{N-1} (t)
\eeq
\label{Markovian_Eqns_of_motion}
\end{subequations}
where, in order to compress notation in our equations, we allowed $i$ to run from 1 to $N-1$ in Eqn. (\ref{Ampl_j_Markovian}), noting that $c_{i-1} (t)=0$ for $i=1$, because qubit $1$ is coupled only to qubit $2$. To simplify the set of Eqns. (\ref{Markovian_Eqns_of_motion}), it is useful to introduce the amplitude transformations $c_i(t) = e^{-i\left[  \omega_e + (N-1) \omega_g \right]t} \Tilde{c}_i(t)$, $i= 1, \dots, N-1$ and $c_N(t) = e^{-i\left[  \omega_e + (N-1) \omega_g -i \gamma_M \right]t} \Tilde{c}_N(t)$, in terms of which the equations become:

\begin{subequations}
\beq
\frac{d \Tilde{c}_i(t)}{dt} = -i \frac{\mathcal{J}}{2} \left[ \Tilde{c}_{i-1} (t) + \Tilde{c}_{i+1} (t) \right], \quad i = 1, \dots, N-1,
\label{Ampl_j_Markovian_tilde}
\eeq
\beq
\frac{d \Tilde{c}_N(t)}{dt}= -i \frac{\mathcal{J}}{2}  e^{\gamma_M t}  \Tilde{c}_{N-1} (t)
\eeq
\label{Markovian_Eqns_of_motion_Tilde}
\end{subequations}
where, as above, $\Tilde{c}_{i-1} (t)=0$ for $i=1$.  The above set of $N$ coupled linear differential equations can be readily solved numerically to yield the time dependence of the tilde amplitudes of the chain sites. The density matrix $\sigma(t)$ of the chain can then be constructed in terms of the site amplitudes as shown in the following equation:

\beq
\sigma(t) = \sum_{i,j}^N c_i(t) c_j^*(t) \ket{\psi_i} \bra{\psi_j}
\label{σ}
\eeq

\subsection{Construction of $ρ(t)$ (non-Markovian system)}
The construction of $ρ(t)$, i.e. the density matrix of the chain with the last qubit interacting with a non-Markovian reservoir, is a  subtler task because the memory effects of the reservoir make the equation of the last qubit amplitude more complicated. In two of our recent papers, we examined the dynamics of XX chain boundary driven by non-Markovian reservoirs \cite{ref16}, as well as the connection between exceptional points and the Quantum Zeno effect in a simpler quantum system consisting of two interacting qubits one of which is coupled to a non-Markovian reservoir \cite{ref50}. On the basis of the approach developed in those papers, as applied to the present system,  we outline here the basic procedure to obtain the amplitudes of the chain sites. In order to discriminate the amplitudes from those of the Markovian damped chain, we denote them by $c'_i (t)$, $i = 1, \dots N $.

Adopting the amplitude transformations $c_i' (t) = e^{-i \left[ \omega_e + (N-1) \omega_g \right] t} {\Tilde{c}'}_i (t)$ , $i = 1, \dots N $ and $ {c'_{\lambda}}^{E}(t) = e^ {-i \left( N\omega_g + \omega_{\lambda} \right)t}  {{\Tilde{c}}_{\lambda}}^{'E}(t)$, the TDSE yields the following equations of motion:
\begin{subequations}
\beq
i \frac{ d \Tilde{c}'_i (t)}{dt} = \frac{\mathcal{J}}{2} \left[ \Tilde{c}'_{i-1} (t) + \Tilde{c}'_{i+1} (t) \right], \quad i = 1, \dots, N-1,
 \label{Amplitudej}
\eeq

\beq
i \frac{ d \Tilde{c}'_N (t)}{dt} = \frac{\mathcal{J}}{2} \Tilde{c}'_{N-1} (t) + \sum_{\lambda} g(\omega_{\lambda}) e^{-i \Delta_{\lambda} t}  \Tilde{c}_{\lambda}^{'E}(t),
 \label{AmplitudeN}
\eeq

\beq
i \frac{ d \Tilde{c}_{\lambda}^{'E}(t)}{dt} = g(\omega_{\lambda}) e^{i \Delta_{\lambda} t} \Tilde{c}'_N (t),
\label{Env1}
\eeq
\end{subequations}
where $\Delta_{\lambda} \equiv \omega_{\lambda} - (\omega_e - \omega_g) \equiv \omega_{\lambda} - \omega_{eg}$ and $\Tilde{c}'_{i-1} (t)=0$ for $i=1$.

Formal integration of Eqn. (\ref{Env1}), under the initial condition $\Tilde{c}_{\lambda}^{'E}(0)=0$,  and substitution into Eqn. (\ref{AmplitudeN}), yields:

\beq
\begin{split}
\frac{ d \Tilde{c}'_N (t)}{d t} = & -i \frac{\mathcal{J}}{2} \Tilde{c}'_{N-1} (t) \\
& - \int_0^t \sum_{\lambda} e^{- i \Delta_{\lambda} \left(  t - t' \right)} \left[  g(\omega_{\lambda}) \right]^2   \Tilde{c}'_N (t')  dt'.
\end{split}
\label{Amplitude_c_N_new}
\eeq

The summation over the environmental modes is at this point replaced by an integration which requires the specification of the environment's spectral density $J(\omega)$, according to the relation $\sum_{\lambda} \left[  g(\omega_{\lambda}) \right]^2 \rightarrow \int d \omega J (\omega)$. The resulting set of differential equations then becomes:  

\begin{subequations}
\beq
\frac{ d \Tilde{c}'_i (t)}{dt} = -i \frac{\mathcal{J}}{2} \left[ \Tilde{c}'_{i-1} (t) + \Tilde{c}'_{i+1} (t) \right], \quad i = 1, \dots, N-1,
\eeq

\beq
\frac{ d \Tilde{c}'_N (t)}{d t}  =  -i  \frac{\mathcal{J}}{2} \Tilde{c}'_{N-1} (t) -  \int_0^t R (t-t') \Tilde{c}'_N (t') dt',
\label{Non-Markovian_Diff_Eqn_Nth_site}
\eeq
\label{Non-Markovian_Diff_Eqns}
\end{subequations}
where 
\beq
R (t) \equiv \int_0^{\infty} J (\omega) e^{- i \left( \omega - \omega_{eg} \right)   t } d\omega.
\label{R(t)}
\eeq
The second term in Eqn. (\ref{Non-Markovian_Diff_Eqn_Nth_site}) reflects the possibility for the excitation to be transferred from the $N^{th}$ qubit to the non-Markovian reservoir and vice versa.

Taking the Laplace transform of Eqns. (\ref{Non-Markovian_Diff_Eqns}) and following the procedure presented in \cite{ref16} we find that the Laplace transform $F_1(s)$ of the tilde amplitude $\Tilde{c}'_1 (t)$ is given by the expression:
\begin{widetext}
\beq
F_1(s) = \frac{ik c_N (0) - k^2 \left[s + B (s) \right] A_1(s) c_{N-1} (0)  + (ik) \sum_{m=1}^{N-2} \Big\{ ik  \left[ s + B (s) \right] A_{N-m}(s) - A_{N-1-m}(s) \Big\} c_m (0)}{ik \left[ s+ B (s) \right] A_N (s) - \left[ 1+  s+ B(s)  \right] A_{N-1} (s)}.
\label{F1}
\eeq
where $k \equiv \frac{2}{\mathcal{J}}$, $B(s)$ is the Laplace transform of the function $R(t)$ and $A_m (s)$ is given by the following expression:
\beq
A_m(s) \equiv \frac{\left[ (iks) +  i \sqrt{k^2 s^2 +4} \right]^m - \left[ (iks) - i \sqrt{k^2 s^2 +4} \right]^m}{2^m i \sqrt{k^2 s^2 +4}},\quad m = 1, \dots, N.
\label{A_m}
\eeq
\end{widetext}
The Laplace transforms $F_i(s)$ of the remaining qubit amplitudes are connected to $F_1(s)$ via the relation:
\beq
\begin{split}
F_i (s) & = A_i(s) F_1 (s) - (ik) \sum_{n=1}^{i-1} A_{i-n}(s) c_n (0), \\
& \quad i = 2, \dots, N,
\label{F_i-F_1}
\end{split}
\eeq
The effectiveness of this method rests on the fact that it allows the derivation of closed form expressions for the Laplace transforms of the tilde amplitudes for an arbitrary number of qubits $N$, an arbitrary spectral density $J(\omega)$ for the environment that interacts with the $N^{th}$ qubit of the chain, as well as arbitrary initial conditions. The Laplace inversion integrals can be readily calculated numerically (or even analytically in some cases) upon specification of these parameters to yield the tilde amplitudes of the chain sites, in terms of which we can calculate $c_i' (t)$ , $i = 1, \dots N $, and thus construct the density matrix for the chain via the relation:

\beq
\rho(t) = \sum_{i,j}^N c'_i(t) {c_j'}^{*}(t) \ket{\psi_i} \bra{\psi_j}
\label{ρ}
\eeq

By construction, the density operators given by Eqs. \eqref{σ} and \eqref{ρ} are Hermitian, with a time-dependence that leads to a decaying trace, for reasons discussed in section II.

Specific types of non-Markovian reservoirs, such as Lorentzian \cite{ref51,ref52}, Lorentzian squared \cite{ref53} and Ohmic \cite{ref54,ref55} are employed in order to assess the effect of non-Markovianity οn the chain in each case. The analytical derivations of the functions $R(t)$ and $B(s)$ for these types of reservoirs are given in Appendix A.

\section{Results $\&$ Discussion}

\begin{figure}[t] 
	\centering
	\includegraphics[width=8.6cm]{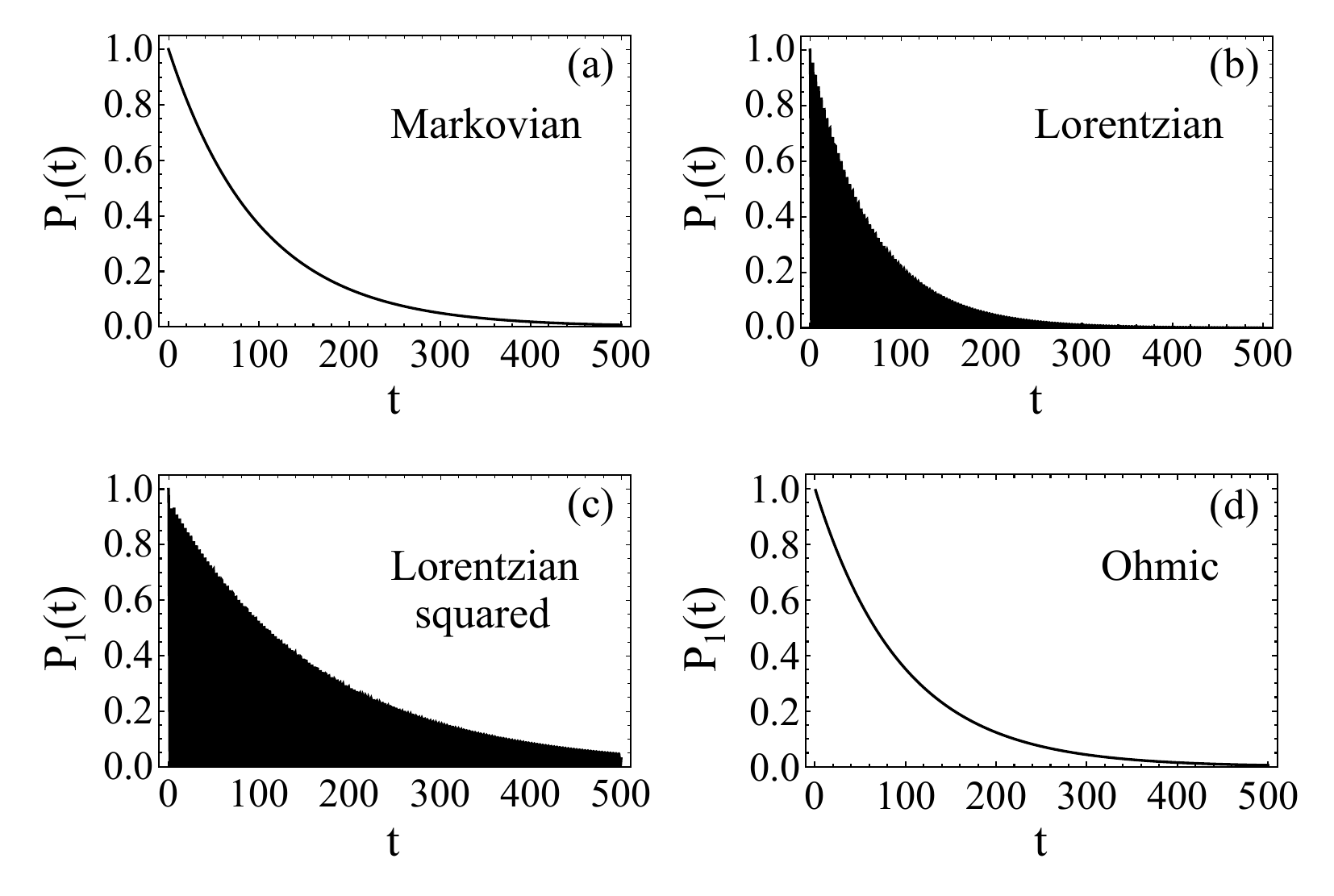}
		\caption[Fig2]{Dynamics of the excitation survival probability of a single qubit coupled to various types of reservoirs: (a) Markovian reservoir with $γ_M =0.01$, (b) Lorentzian reservoir with $g=1$, $γ=0.03$ and $Δ_c=0$, (c) Lorentzian squared reservoir with $g=1$, $γ=0.3$ and $Δ_c=0$, (d) Ohmic reservoir with $g=1$, $S=1.5$, $ω_c=8$ and qubit frequency $ω_{eg}=10$.  }
		\label{Fig2}
\end{figure}

\begin{figure*}[t] 
	\centering
	\includegraphics[width=17.6cm]{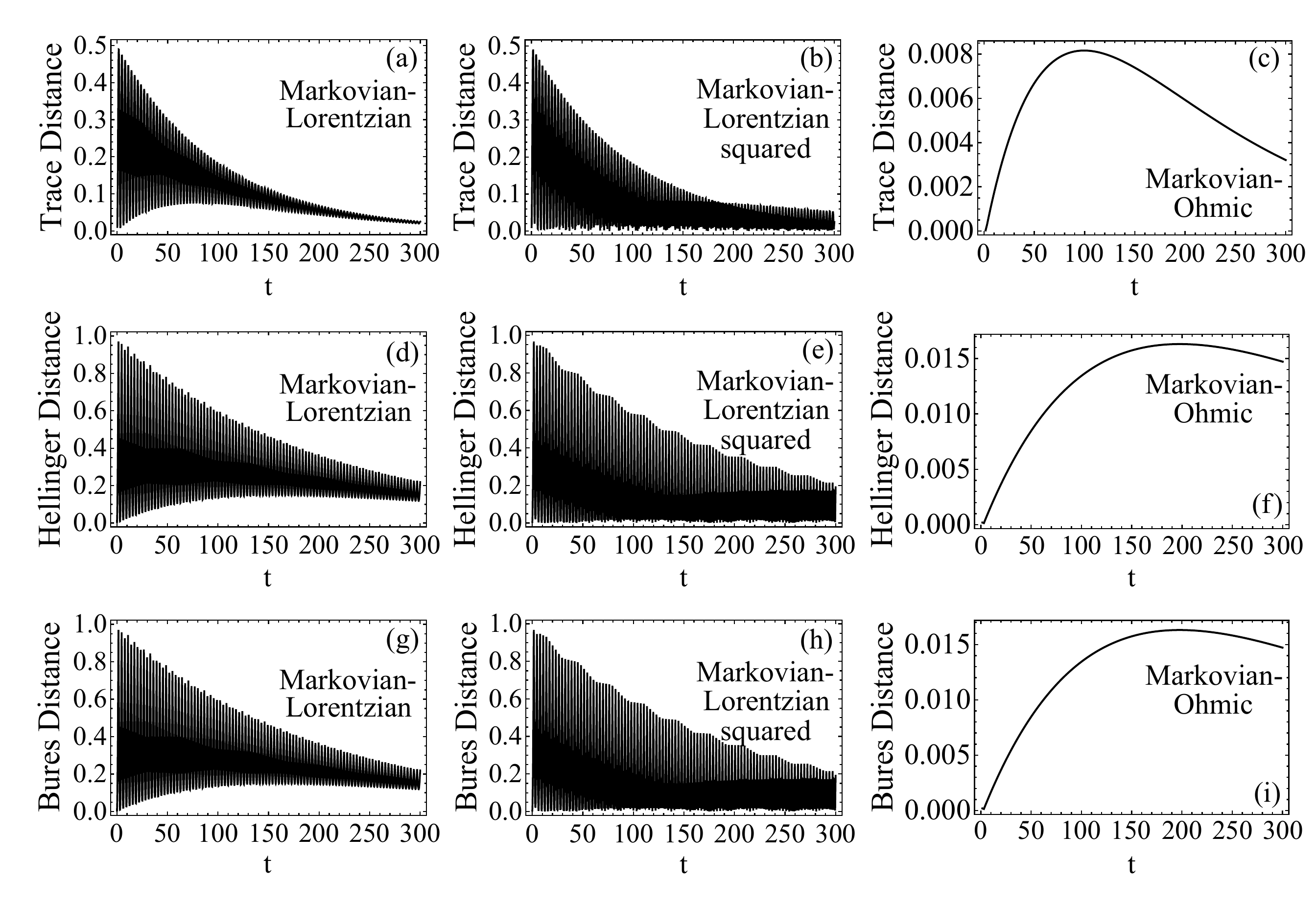}
		\caption[Fig3]{QSD measures between the Markovian-damped and the non-Markovian-damped system using various types of non-Markovian reservoirs in the single-qubit case. The types of reservoirs chosen and their parameters are: Markovian reservoir with $γ_M =0.01$,  Lorentzian reservoir with $g=1$, $γ=0.03$ and $Δ_c=0$, Lorentzian squared reservoir with $g=1$, $γ=0.3$ and $Δ_c=0$, Ohmic reservoir with $g=1$, $S=1.5$, $ω_c=8$ and qubit frequency $ω_{eg}=10$. (a) Markovian-Lorentzian Trace Distance, (b) Markovian-Lorentzian-squared Trace Distance, (c) Markovian-Ohmic Trace Distance, (d) Markovian-Lorentzian Hellinger Distance, (e) Markovian-Lorentzian-squared Hellinger Distance, (f) Markovian-Ohmic Hellinger Distance, (g) Markovian-Lorentzian Bures Distance, (h) Markovian-Lorentzian-squared Bures Distance, (i) Markovian-Ohmic Bures Distance.}
		\label{Fig3}
\end{figure*}

The stated objective of this paper is the quantitative assessment of NM of non-Markovian reservoirs in a realistic context provided by a spin chain. As outlined in the previous sections, we have chosen to assess NM in terms of QSD measures between the state of the chain coupled to various reservoirs. An obviously suitable for the purpose quantity is the QSD between the density operator of the chain evolved under a Markovian from that evolved under a non-Markovian reservoir. From a physical standpoint,  it stands to reason that the larger the values of the QSD, irrespective of the measure under consideration, the less Markovian the system's evolution will be, which means more pronounced non-Markovianity. Given that reservoirs by definition involve certain parameters, characterising their properties, in addition to the system-reservoir coupling constant, a decision has to be made as to the choice of parameters that make the value of QSD most meaningful. It is rather obvious that a direct comparison between, for example, an evolution under a Markovian reservoir with damping rate $\gamma_M$ and a Lorentzian one with arbitrarily chosen parameters $g$, $\gamma$ and $\Delta_c$ would be pointless. We suggest that, for the comparison to be meaningful, a feature common to the effect of all reservoirs entering the comparison should be adopted. With that in mind, we note that for all reservoirs the excitation of all qubits will eventually decay to zero. Since initially the excitation is in the first qubit, without claim to uniqueness, we have chosen the parameters of the various reservoirs such that the half life of the excitation of the first qubit is approximately the same to its half life under a Markovian reservoir with a given $\gamma_M$. This convention leaves a freedom on the choice of the parameter $g$, which expresses the coupling strength between the last qubit of the chain and the non-Markovian reservoir, and consequently it regulates the frequency of population exchange between the two. In both the single-qubit as well as the many-qubit cases, and for all non-Markovian reservoirs considered, we fix the parameter $g$ at the value $g=1$. Note that the parameters that determine the lifetime of the excitation in the chain are chosen such that the period of population oscillations between the last qubit and the reservoir is smaller than this lifetime.

\subsection{Single qubit}
We begin our analysis by considering the simplest case of a single qubit coupled to a variety of non-Markovian reservoirs. These  results serve as a point of reference in the comparative analysis of the interplay between non-Markovianity  and qubit-qubit interaction, as reflected in the evolution of the QSD measures as a function of the time, analysed in section IVB.

In Fig. 2 we plot the time dynamics of the single qubit excitation survival probability for various types of reservoirs. In particular, we examine the cases of Markovian, Lorentzian, Lorentzian squared and Ohmic reservoirs, with the parameters of each non-Markovian reservoir chosen as described above. The Lorentzian and Lorentzian squared spectral densities result in general to fast oscillations in the single-qubit survival of excitation dynamics, indicating the exchange of the excitation between the qubit and the environment within finite times. The high frequency of these oscillations is attributed to a large coupling strength $g$ compared to the Lorentzian and Lorentzian-squared widths $\gamma$. As $g$ is decreased the frequency of oscillations tends also to decrease. Note that the Markovian limit would be captured by increasing $\gamma$ but at the same time keeping the ratio $g^2/\gamma$ constant. The Ohmic spectral distribution tends in general to be much broader than the Lorentzian and Lorentzian squared distributions, leading to a qubit excitation survival probability that resembles more the respective survival probability of the excitation for the Markovian case.

The QSD measures between the Markovian-damped single-qubit system and the non-Markovian ones are presented in Fig. 3. In all three of the non-Markovian reservoirs under consideration, the coupling strength $g$ between the single-qubit and the reservoir has been kept the same. As expected, all QSD measures are zero for $t=0$, as well as in the long time limit, as  there is practically no excitation left in the open system. 

We focus first on the comparison  of the degrees of non-Markovianity, for different types of non-Markovian reservoirs, resulting from the same QSD measure. The trace distance between the Markovian and the Lorentzian damped system (Fig. 3(a)) exhibits rapid oscillations between two bounds, whose values change over time. The upper bound is initially at its maximum value and over time decreases to zero, whereas the lower bound initially increases, reaches a maximum, after which it decreases merging with the upper bound,  tending eventually to zero for long times. The  dynamics of the trace distance between the Markovian and the Lorentzian-squared damped system (Fig. 3(b)) also exhibit rapid oscillations, following a trend similar to the Markovian-Lorentzian case, with the exception that the lower bound can be zero within finite times, indicating that, at those instants, the Lorentzian-squared damped system resembles the Markovian-damped one. The oscillations in both Figs. 3(a) and 3(b) are indicative of the fast excitation exchange between the qubit and the non-Markovian reservoirs. In Fig. 3(c) we plot the trace distance between the Markovian-damped and the Ohmic-damped qubit state as a function of  time. Clearly, the behaviour of the trace distance dynamics in this case differs substantially from the dynamics of the Markovian-Lorentzian and the Markovian-Lorentzian-squared cases, in that it does not exhibit any oscillations, reaches a maximum after some finite time before decreasing, with the overall values of the trace distance remaining much smaller than the respective values in panels (a) and (b) of Fig. 3. We may therefore safely conclude that, under the prescribed conditions on parameters, the Ohmic-damped single qubit system displays the lowest character of NM.

In addition to the trace distance, we have calculated QSD's in terms of two other relevant measures, namely the Hellinger and Bures measures, with the results shown in Fig. 3. The overall behavior bears significant similarity to that of the trace distance, with some non-negligible quantitative differences. Note that all Markovian-Lorentzian and Markovian-Lorentzian-squared QSD measures tend to zero as we tune the Lorentzian and Lorentzian-squared parameters to the Markovian limit. In Fig. 3, the oscillations reflecting the exchange of excitation between qubit and reservoir are still present in both measures for the Markovian-Lorentzian and Markovian-Lorentzian-squared cases, whereas they are absent for the Markovian-Ohmic one. However both of these measures produce generally slower decrease of the QSD from the respective trace distance measure. For example, in terms of the trace distance, the value of the QSD between Lorentzian and Markovian in the vicinity of $t = 100$ hovers around $0.15$, whereas the Hellinger and Bures measures indicate values around a mean of $0.4$. Moreover, the oscillations of the QSD for the Markovian obtained from these two measures are damped much more slowly than the oscillations in the trace distance. The Ohmic case seems to follow the same trend, in that these two measures produce larger values for the QSD and a much slower evolution towards the expected value of $0$. In summary, for all three non-Markovian reservoirs in the single qubit case, the trace distance is found to show faster decay of non-Markovianity than the Hellinger and Bures measures do. It is also worth noting that, despite the non-trivial formal difference of the expressions for the Hellinger and Bures QSD (Eqns. (\ref{Hellinger_open}) and (\ref{Bures_open})), they lead to practically the same dynamics, indicating the same NM effect.  

\subsection{Five mutually interacting qubits} 

\begin{figure}[t] 
	\centering
	\includegraphics[width=8.6cm]{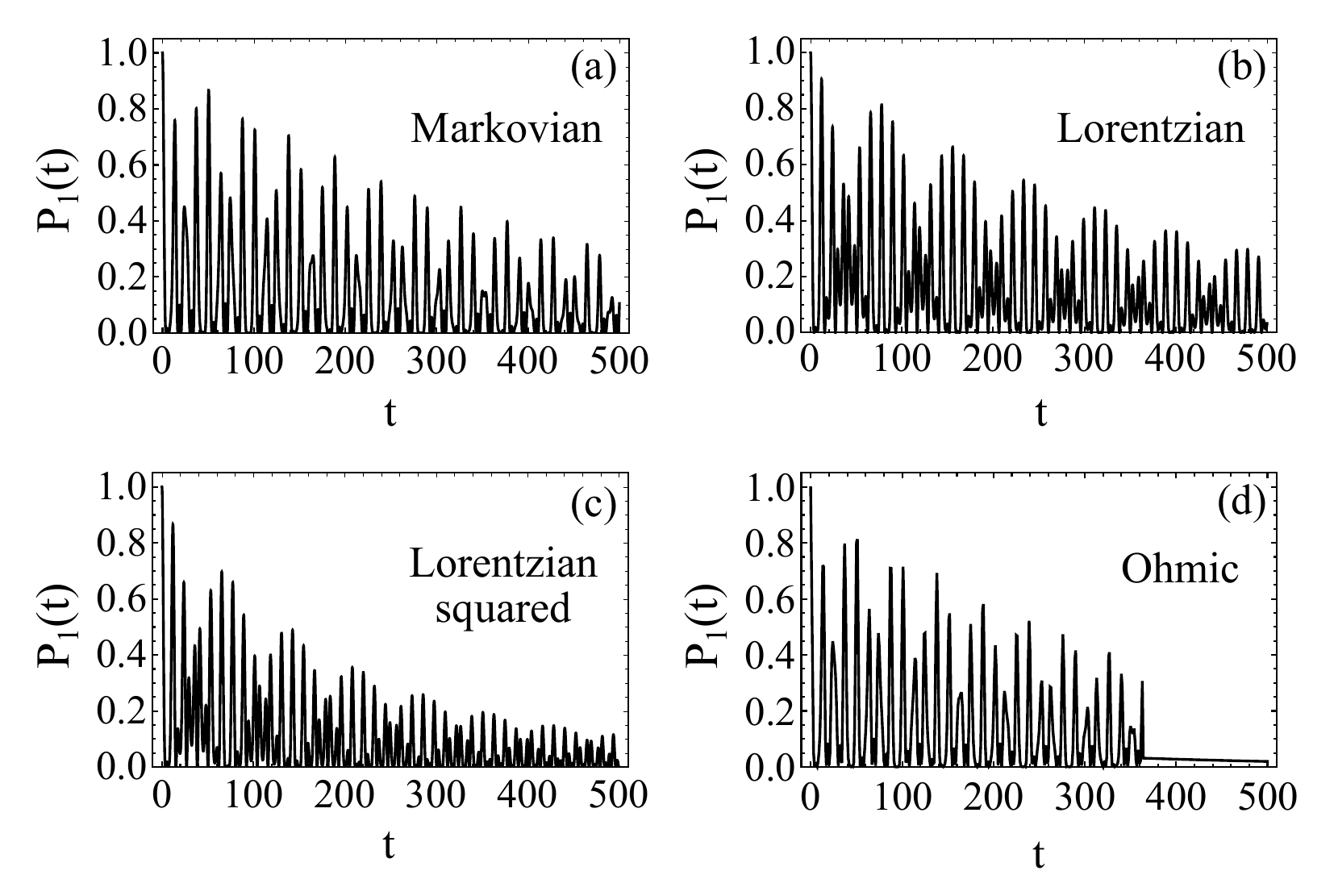}
		\caption[Fig4]{Dynamics of the excitation survival probability of the first qubit of a chain consisting of $N=5$ qubits with qubit-qubit coupling $\mathcal{J}=1$. The initial excitation is chosen to be on the first qubit while the last qubit of the chain is coupled to various types of reservoirs: (a) Markovian reservoir with $γ_M =0.01$, (b) Lorentzian reservoir with $g=1$, $γ=0.03$ and $Δ_c=0$, (c) Lorentzian squared reservoir with $g=1$, $γ=0.3$ and $Δ_c=0$, (d) Ohmic reservoir with $g=1$, $S=1.5$, $ω_c=8$ and qubit frequency $ω_{eg}=10$.}
		\label{Fig4}
\end{figure}

\begin{figure*}[t] 
	\centering
	\includegraphics[width=17.6cm]{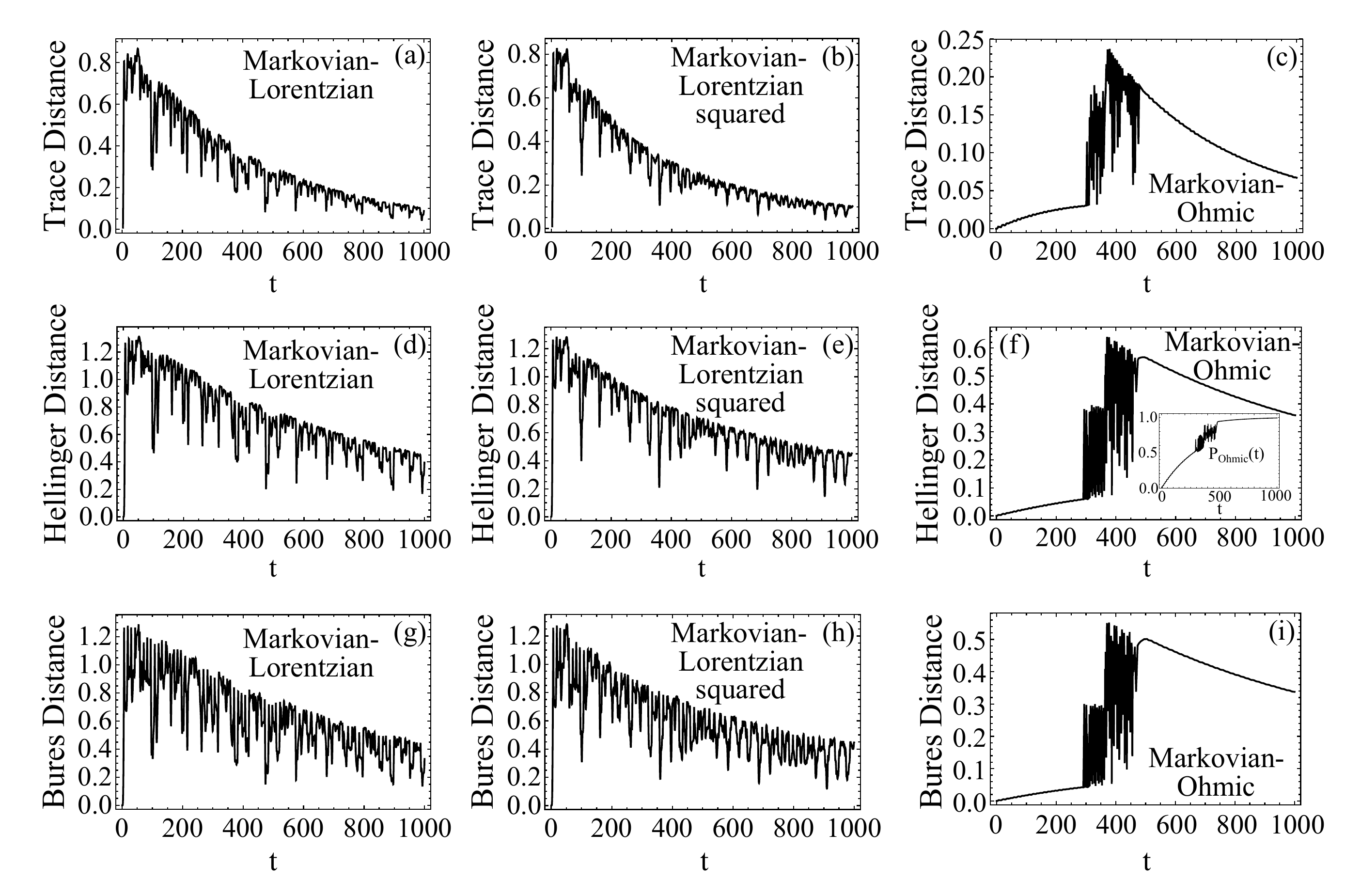}
		\caption[Fig5]{QSD measures between the Markovian-damped and the non-Markovian-damped system using various types of non-Markovian reservoirs for the case of $N=5$ qubits with qubit-qubit coupling $\mathcal{J}=1$. The types of reservoirs chosen and their parameters are: Markovian reservoir with $γ_M =0.01$,  Lorentzian reservoir with $g=1$, $γ=0.03$ and $Δ_c=0$, Lorentzian squared reservoir with $g=1$, $γ=0.3$ and $Δ_c=0$, Ohmic reservoir with $g=1$, $S=1.5$, $ω_c=8$ and qubit frequency $ω_{eg}=10$. (a) Markovian-Lorentzian Trace Distance, (b) Markovian-Lorentzian-squared Trace Distance, (c) Markovian-Ohmic Trace Distance, (d) Markovian-Lorentzian Hellinger Distance, (e) Markovian-Lorentzian-squared Hellinger Distance, (f) Markovian-Ohmic Hellinger Distance, (g) Markovian-Lorentzian Bures Distance, (h) Markovian-Lorentzian-squared Bures Distance, (i) Markovian-Ohmic Bures Distance. In the inset of panel (f) we show the dynamics of the probability of finding the excitation in the Ohmic environment for the case of $N=5$ qubits. The time window where the probability dynamics abruptly become oscillatory coincides with the time window where the Markovian-Ohmic QSD measures follow the same behaviour.}
		\label{Fig5}
\end{figure*}

The merit of our approach rests with the possibility to obtain straightforwardly the Laplace transforms of the qubits tilde state amplitudes for an arbitrary number of qubits $N$, because $N$ appears in our equations as a modifiable parameter. Having calibrated the measures of QSD in characterising non-Markovianity in the simplest set-up of a single qubit, in this section we explore the role of non-Markovianity in the more realistic situation of a chain involving qubit-qubit interaction. For a concrete quantitative analysis, we have chosen five qubits in an XX chain with nearest neighbor coupling $\mathcal{J}$. The calculation follows the approach presented in section III, with the initial excitation in qubit $1$ and the last qubit $5$ coupled to the reservoir. The question now is how the non-Markovianity of the reservoir affects the exchange of excitation between  the qubits, as manifested in the evolution of QSD between the Markovian and the various non-Markovian reservoirs.

In a set-up of this type, the crucial parameter is $\mathcal{J}$, as it controls the communication between qubits, which typically is the dominant aspect in applications. In order to remain within a realistic scenario, we have chosen the Markovian damping rate $\gamma_M$ much smaller than $\mathcal{J}$, keeping the choice of the parameters of the non-Markovian reservoirs as in the previous section, so that the initial excitation of the first qubit decays on a time scale approximately the same for all reservoirs. At first,  we calculate the time evolution of the excitation probability of the first qubit for all of the reservoirs under consideration,  with the results shown in Fig. 4. The glaring difference between the single-qubit dynamics in Fig. 2 and the $N=5$ dynamics of the first qubit excitation in  Fig. 4, is the appearance of oscillations in the dynamics of the latter, indicating the exchange of population between the qubits of the chain. In the cases of Lorentzian and Lorentzian squared  reservoirs, such oscillations are superimposed on the oscillations between the last qubit and the reservoir, resulting to the characteristic oscillatory dynamics of $P_1(t)$ depicted in Figs. 4(b) and 4(c). For the Markovian case  (Fig. 4(a)) the oscillations are due solely to the coupling between the qubits, because as we have seen in the case of the single qubit (Fig. 2(a)) the Markovian damping is monotonic. Again, for the Ohmic-damped case, the dynamics of the first qubit population (Fig. 4(d)) bears strong similarity to the Markovian-damped chain, with a rather unexpected exception, in that the oscillations are found to collapse abruptly after a finite time, a behaviour that has been also reported in previous works involving qubits coupled to Ohmic reservoirs \cite{ref16,ref56}.

We consider now the dynamics of the QSD measures exploring the non-Markovianity of the three reservoirs in the context of a chain of $N=5$ interacting qubits, with the results shown in Fig. 5. First we note that the time scales of decay for all QSD measures are longer than the respective time scales in the single-qubit case of Fig. 3. This behaviour was to be expected, since the excitation is now spread over the whole chain and naturally it takes more time to be lost in the reservoir. At the same time, the values of the QSD measures are in general larger than the respective values of Fig. 3, indicating that the NM of the open system is affected by the number of qubits in the chain. The Markovian-Lorentzian and Markovian-Lorentzian-squared QSD measures are seen to display an overall similar behaviour. This similarity requires some clarification, as it is a bit misleading. Recall that the values of the parameters for the Lorentzian and Lorentzian-squared spectral densities have been chosen according to the criterion explained in the discussion of Fig. 1. As a consequence, the comparison is not between a Lorentzian and its square. However, in detailed calculations we have found that an open system damped by a Lorentzian squared reservoir would display  much larger NM character compared to the same system damped by a Lorentzian with the same parameters ($g$, $γ$ and $Δ_c$). In fact, under those conditions, the excitation throughout the chain for the Lorentzian squared case remains constant for a time so long that it practically resembles a steady state. Further discussion of this feature is presented in section  V. All QSD measures we examined (Trace distance, Hellinger distance and Bures distance) exhibit dynamics following the same trend, although the exact distance values may differ in general between different measures. As clearly seen in Fig. 5, the Hellinger and Bures distances display minor differences in their dynamical behaviour for all of the reservoirs studied.

A rather interesting phenomenon appears in the dynamical behaviour of the QSD measures in the case of the Ohmic-damped chain. In contrast to the single-qubit case, the interplay between qubit-qubit and qubit- Ohmic reservoir interactions gives rise to an intricate behavior in the dynamics for all of the Markovian-Ohmic QSD measures considered. The dynamics are initially smooth for all QSD measures with the distance being zero at $t=0$ and slowly increasing afterwards. However, there is a time when they suddenly change and exhibit rapid oscillations. These rapid oscillations appear only in a time window which is the same for all QSD measures considered. Beyond that time window, the dynamics are smoothed out with the distance slowly decaying back to zero. The time window of that sudden modification of the behaviour of the QSD measures coincides with the time window within which the total probability of finding the excitation in the Ohmic reservoir, exhibits similar behaviour (see inset of Fig. 5(f)). Changes in the dynamical behaviour of the system can be also observed through abrupt changes in the dynamics of the qubit excitations of the chain, some of which were reported in one of our previous papers \cite{ref16}. In summary, the qubit-qubit interaction is found to enhance the NM obtained in all QSD measures. Conversely, this may be interpreted as an enhancement of the excitation storage in the chain due to the non-Markovian character of the reservoir.

\section{Summary and Outlook}

The work in this paper was undertaken with the purpose of evaluating the NM of some typical non-Markovian reservoirs, by means of the most common QSD measures, some of which required a slight modification owing to the character of the density operators involved. In order to offer an analysis connected with a realistic system typical in quantum information processing, we chose an XX chain of interacting qubits with a reservoir connected to one end of the chain. QSD measures require the density matrices of the systems under consideration. Since non-Markovian reservoirs do not in general lend themselves to descriptions in terms of a master equation, we employed an approach developed in earlier work \cite{ref16} based on the solution of the Schr{\"o}dinger equation, in which the amplitudes of the excitation of the sites as a function of time are obtained by means of Laplace transform. With those amplitudes in hand we construct the density matrix of the system which is necessary for the calculation of the desired QSD measure.  

As a measure of NM we have chosen the distance of the quantum state of our system as evolved under a given non-Markovian reservoir from the state evolved under a Markovian one. A meaningful comparison required the judicious choice of the parameters entering the expressions of the spectral densities of the reservoirs involved in each QSD calculation. Given the flexibility of our approach as to the number of qubits it can handle, we studied first the case of a single-qubit coupled to a reservoir, so as to have a frame for the evaluation of the interplay between qubit-qubit interaction and qubit-reservoir. The results from the single qubit analysis showed that the Hellinger and Bures measures indicated slightly higher character of NM for all three reservoirs than the trace distance measure did. On the other hand, all three measures have indicated significantly lower character of NM for the Ohmic reservoir, in comparison to the Lorentzian or Lorentzian squared. Moreover, the absence of any noticeable back-flow of excitation from the Ohmic reservoir to the qubit suggests that the Ohmic reservoir behaves practically in a Markovian fashion. 

As noted in section IV, the Lorentzian and Lorentzian squared reservoirs, on the basis of all three measures, were found to exhibit essentially the same character of NM. This might seem counter intuitive, because inspection of the relevant formal expressions in the appendix suggests that the squared Lorentzian is more peaked than the Lorentzian with the same parameters, which would imply a higher character of NM. However, the Lorentzian squared employed in all of our comparative calculations (Figs. 2-5) was not the square of the accompanying Lorentzian, owing to the adjustment of the parameters explained in section III. If we calculate the evolution of the qubit excitation dynamics and evaluate the character of NM for a Lorentzian and a Lorentzian squared spectral density characterized by the same parameters ($g$, $γ$ and $Δ_c$), we do indeed find more back-flow and higher NM character for the Lorentzian squared, which brings up an interesting issue in reservoir engineering. If in any open system subjected to dissipation through a Lorentzian reservoir, it were possible to modify the reservoir to the square of the Lorentzian, dissipation would slow down significantly. Our consideration of the Lorentzian squared spectral density is not a mere mathematical exercise. It was inspired by the observation of non-Lorentzian line shapes \cite{ref53,ref57}, including the Lorentzian squared, albeit in different physical context.  

The results for the case of five interacting qubits chain bear similarities to as well as differences from the single-qubit case. The similarity is seen in the NM by the three QSD measures which produce more or less similar values, with the value of the trace distance somewhat lower than the values of the Hellinger and Bures measures. In other words, the three measures provide mutually compatible results. However, the measures of the NM for all three reservoirs for the chain are noticeably higher than those for the single-qubit. Moreover, for the Ohmic reservoir, the QSD obtained by all three measures is not only quantitatively but even qualitatively different from that of the single qubit. The smooth dynamics obtained in the single qubit case, is in this case interrupted by a region of abrupt oscillations, within a certain time window in which the dynamics of the probability of finding the excitation in the Ohmic environment was found to exhibit the same behaviour. However, the stunning difference between the single and five qubit cases lies in the retention of population distributed among the interacting qubits and the much slower decay of the excitation for all of the reservoirs in the latter case, which is compatible with the consistently higher measures of NM reported in our work. It can then be argued that the exchange of excitation between the qubits slows down the dissipation which implies that the larger the number of the qubit the slower the loss into the reservoir. Perhaps the most significant message conveyed by the comparison of the single-qubit and chain of interacting qubits results is that the NM is not independent of the quantum system in the context of which it is evaluated.

In closing, we believe that the appropriate modifications of the QSD measures, that extend their use to non-trace-preserving processes and the detailed analysis of their applicability to a qubit chain system, have revealed subtle aspects of non-Markovianity. Extension of our approach and analysis to other open systems can be expected to offer new insights into the performance of protocols pertaining to quantum information technology tasks as well as to the statistical mechanics of spin chains. Even for the single chain coupled to non-Markovian reservoirs, issues such as the dynamics of entanglement, and the NM beyond single excitation represent potentially fruitful territory for exploration. 

\section*{Acknowledgments}
GM would like to acknowledge the Hellenic Foundation for Research and Innovation (HFRI) for financially supporting this work under the 3rd Call for HFRI PhD Fellowships (Fellowship Number: 5525).

\begin{figure}[H] 
	\centering
	\includegraphics[width=4cm]{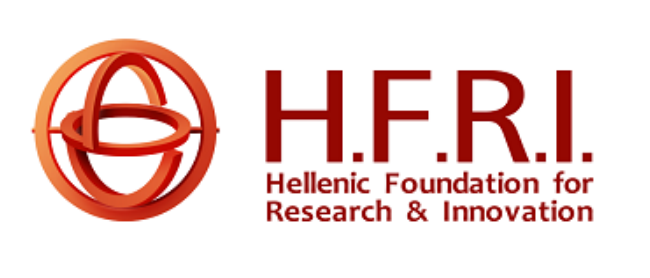}
\end{figure}

\appendix
\renewcommand{\thesection}{\Alph{section}}
\numberwithin{equation}{section}

\section{Derivations of $R(t)$ and $B(s)$ for special types of non-Markovian reservoirs}

In this appendix we provide the analytical derivations of the functions $R (t) \equiv \int_0^{\infty} J (\omega) e^{- i \left( \omega - \omega_{eg} \right)   t } d\omega$ and their Laplace transforms $B(s)$ for various forms of non-Markovian spectral densities $J(\omega)$. The calculation of $B(s)$ in each case is necessary in order to invert the Laplace transforms $F_i(s)$, $i = 1, \dots, N$ and find the expressions of the tilde amplitudes of the chain.

\subsection{Lorentzian spectral density}
For reservoirs characterized by Lorentzian spectral densities, $J(\omega)$ is given by:
\beq
J (\omega) = \frac{g^2}{ \pi} \frac{\frac{\gamma}{2}}{\left( \omega-\omega_c \right)^2 + (\frac{\gamma}{2})^2},
\eeq
where $g$ is the coupling strength between the last qubit of the chain and the reservoir, while $\gamma$ and $\omega_c$ is the width and the peak of the distribution, respectively.

An approximation that considerably simplifies analytically the calculation of the function $R(t)$ is the extension of the lower limit of the integration over frequency from 0 to $- \infty$. Note that such extension is not in general valid for any spectral density; it is however well justified in the case of a Lorentzian with positive peak frequency and width such that the distribution has practically negligible extension to negative frequencies. The necessary condition for this approximation is $\gamma$ $<<$ $ \omega_c$. In this case the frequency integral can be calculated analytically, yielding:
\beq
R(t) = g^2 e^{- i \Delta_c t} e^{-\frac{\gamma}{2} t},
\label{R(t)_Lorentzian}
\eeq
where $\Delta_c \equiv \omega_c - \omega_{eg}$. The Laplace transforms of $R(t)$ is then given by the expression:

\beq
B (s) = \frac{g^2}{s+ \frac{\gamma}{2} + i \Delta_c}.
\eeq

\subsection{Lorentzian squared spectral density}
The Lorentzian squared spectral density is a special case of a broader set of spectral densities referred as powers of Lorentzian that are given by the form:
\beq
J(\omega) = g^2 \mathcal{N}_n  \frac{\left(\frac{\gamma}{2}\right)^{2n-1}}{\left[ \left( \omega-\omega_c \right)^2 + (\frac{\gamma}{2})^2 \right]^n},
\eeq
where $\mathcal{N}_n$ is an appropriate normalization factor that depends upon the non-zero integer $n$. For $n=1$ we capture the case of the Lorentzian spectral density, while for $n=2$ we capture the Lorentzian squared spectral density given by:

\beq
J(\omega) = \frac{2g^2}{\pi} \frac{\left(\frac{\gamma}{2}\right)^{3}}{\left[ \left( \omega-\omega_c \right)^2 + (\frac{\gamma}{2})^2 \right]^2},
\eeq

Following the same procedure as with the case of the Lorentzian, we find that the function $R(t)$ is given by:

\beq
R(t) = g^2 \left( 1 + \frac{\gamma t}{2}  \right) e^{- i \Delta_c t} e^{-\frac{\gamma}{2} t}.
\label{R(t)_Lorentzian_squared}
\eeq
Notice that Eqn. (\ref{R(t)_Lorentzian_squared}) is the same as Eqn. (\ref{R(t)_Lorentzian}) with an addition of a multiplication factor $\left( 1 + \frac{\gamma t}{2} \right)$. The Laplace transform of Eqn. (\ref{R(t)_Lorentzian_squared}) reads:

\beq
B(s) = g^2 \frac{\left( s + \gamma + i \Delta_c \right)}{\left( s + \frac{\gamma}{2} + i \Delta_c \right)^2}
\eeq

\subsection{Ohmic spectral density}
The Ohmic spectral density is given by the expression:

\beq
J(\omega) = \mathcal{N} g^2 \omega_c \left( \frac{\omega}{\omega_c} \right)^{\mathcal{S}} \exp \left(- \frac{\omega}{\omega_c} \right),
\label{OhmicSD}
\eeq
where $\mathcal{S}$ is the Ohmic parameter, characterizing whether the spectrum of the reservoir is sub-Ohmic ($\mathcal{S} < 1$), Ohmic ($\mathcal{S}$ = 1) or super-Ohmic ($\mathcal{S} > 1$), $\omega_c$ is the Ohmic cut-off frequency and $\mathcal{N}$ is a normalization constant given by the relation $\mathcal{N}= \frac{1}{ \left( {\omega_c} \right)^2 \Gamma \left( 1 + \mathcal{S} \right)}$, where $\Gamma(z)$ is the gamma function. 

Substituting Eqn. (\ref{OhmicSD}) back to the expression of $R(t)$ and calculation of the frequency integral yields:

\beq
R(t) = g^2  e^{i \omega_{eg} t} \left( i \omega_c t + 1 \right)^{-1-\mathcal{S}}
\eeq
The Laplace transform of $R(t)$ can then be calculated analytically, yielding:
\beq
B(s) = - g^2 \frac{i^{1- \mathcal{S}}}{\omega_c}   e^{-i K(s)}  \left[ K(s) \right]^{\mathcal{S}} \Gamma \left(- \mathcal{S}, - i K(s) \right), 
\label{B_Ohmic}
\eeq
where $K(s) \equiv \left( s-i \omega_{eg} \right) / \omega_c$ and $\Gamma(a,z)$ is the incomplete gamma function.

\end{document}